\documentclass[reprint,
superscriptaddress,
 amsmath,amssymb,
 aps,
 pra,
]{revtex4-1}

\usepackage{graphicx}
\usepackage{dcolumn}
\usepackage{bm}
\usepackage{hyperref}

\begin{document}
\title{Quasinormal-mode modeling and design in nonlinear nano-optics}

\author{Carlo Gigli}
\affiliation{Matériaux et Phénomènes Quantiques, Université de Paris \& CNRS, 10 rue A. Domon et L. Duquet, 75013 Paris, France}
\author{Tong Wu}%
\affiliation{LP2N, Institut d’Optique Graduate School, CNRS, Univ. Bordeaux, 33400 Talence, France }
\author{Giuseppe Marino}
\affiliation{Matériaux et Phénomènes Quantiques, Université de Paris \& CNRS, 10 rue A. Domon et L. Duquet, 75013 Paris, France}
\author{Adrien Borne}
\affiliation{Matériaux et Phénomènes Quantiques, Université de Paris \& CNRS, 10 rue A. Domon et L. Duquet, 75013 Paris, France}
\author{Giuseppe Leo}
\affiliation{Matériaux et Phénomènes Quantiques, Université de Paris \& CNRS, 10 rue A. Domon et L. Duquet, 75013 Paris, France}
\author{Philippe Lalanne}
\affiliation{LP2N, Institut d’Optique Graduate School, CNRS, Univ. Bordeaux, 33400 Talence, France }

\date{\today}

\begin{abstract}
	Based on quasinormal-mode theory, we propose a novel approach enabling a deep analytical insight into the multi-parameter design and optimization of nonlinear photonic structures at subwavelength scale. A key distinction of our method from previous formulations relying on multipolar Mie-scattering expansions is that it directly exploits the natural resonant modes of the nanostructures, which provide the field enhancement to achieve significant nonlinear efficiency. Thanks to closed-form expression for the nonlinear overlap integral between the interacting modes, we illustrate the potential of our method with a two-order-of-magnitude boost of second harmonic generation in a semiconductor nanostructure, by engineering both the sign of $\chi^{(2)}$ at subwavelength scale and the structure of the pump beam.
\end{abstract}

	\maketitle
	
	\section{Introduction}

	Nonlinear optical processes mediated by second-, third-, or higher-order nonlinearities play a crucial role in many photonic applications, including ultrashort-pulse shaping \cite{DeLong1994,Arbore1997}, spectroscopy \cite{Heinz1982}, generation of novel states of light \cite{Kuo2006, Krischek2010}, and quantum information processing \cite{Tanzilli2005}. Because $\chi^{(2)}$  and $\chi^{(3)}$ are generally weak, a well-known approach for lowering the power requirements of devices is to enhance nonlinear interactions by employing optical resonances. 
	While high nonlinear efficiencies have been reported in cavities with large quality factor Q and wavelength-scale volume \cite{Notomi2010}, in recent years there has been significant interest in their counterparts at the nanoscale, where both metallic \cite{Kauranen2012} and dielectric particles supporting small-Q Mie resonances \cite{Smirnova2016} have been explored with two aims: 1) reduce the size of nonlinear components towards functional nanophotonic circuitry; and 2) lower their response time, allowing the manipulation of optical signals at femtosecond scale. 
	In the case of plasmonic resonators, where the electromagnetic field is tightly confined close to the surface and intrinsic absorption losses are huge, second harmonic generation (SHG) efficiency $\eta_{SHG}\sim 5\cdot 10^{-10}W^{-1}$ has been reported \cite{Celebrano2015}. Interestingly, the tunability of plasmonic modes has also been exploited to shape the resonator response for nonlinear holography \cite{Almeida2016}.  On the other hand, high-contrast dielectric nanoparticles exhibit light confinement inside their volume, enabling to exploit the bulk properties of the material to boost the nonlinear response. This firstly motivated the study of third-order processes, with applications ranging from beam shaping to optical switching \cite{Shcherbakov2015}. The same advantage was then exploited in non-centrosymmetric materials, with $\eta_{SHG}\sim 6\cdot 10^{-6}  W^{-1}$ \cite{Gili2016,Camacho-Morales2016}. The number of related studies is becoming relevant and new applications continuously emerge, yet a robust and unified modal theory seems to be missing for sub-wavelength nonlinear optics.
	
	Currently, the design of nanoresonators with tailored nonlinear responses is a complex task due the presence of several resonances at each harmonic frequency, and the complexity in matching the driving field and the resonator modes. Most designs rely on brute force computations, are rarely coupled to optimization procedures \cite{Hughes2019},  and are in all cases computationally involved and CPU demanding. They are also inconveniently interpreted with multipolar Mie expansions \cite{Smirnova2016, Kivshar2018, Shcherbakov2015a, Kruk2017}. While Mie formalism is simple and powerful for studying the scattering properties of spherical particles suspended in a uniform medium \cite{Frizyuk2019}, it is no longer analytical for more complex geometries or particles on substrates. This inevitably leads to a loss of computational efficiency and physical insight. Additional difficulties arise in the case of multipolar decomposition of non-spherical nanoparticles, since the decomposition varies with the frequency and incidence angle of the driving field. While approximate solutions have been proposed in literature, like the field decomposition inside a finite-length cylindrical resonator over the complete set of modes of the corresponding infinitely long cylinder \cite{Guasoni2017}, they are not of general usage.
	
	In this context, a theory based on the resonant modes appears more appropriate and natural to adopt, as it is commonly the case for nonlinear processes in waveguides and photonic crystals \cite{Berger1998}, because it promotes important concepts such as mode overlap, phase matching and field enhancement. At variance with closed resonators, once excited, these open cavities modes exponentially decay in time. The modes of such non-Hermitian problems are referred to as quasinormal modes (QNMs) and are mathematically found as time-harmonic solutions of source-free Maxwell’s equations \cite{Lalanne2018}. Due to their non-conservative nature, QNMs exhibit complex eigenfrequencies, denoted by $\tilde{\omega}_m$ in the following. Theoretical QNM formalisms have been initially established for simple and compact resonator geometries (e.g. 1D Fabry-Perot cavities, Mie sphere resonators \cite{More1971, Leung1994, Doost2014, Colom2018}) in a uniform background, for which analytical expressions of the field are available. It is only recently that complex resonators with different shapes, made of dispersive materials with several possible inclusions (like plasmonic oligomers) or possibly placed in complex environments (e.g. deposited on a substrate) have been analyzed with QNM theory. This progress was enabled by: 1) the normalization of QNM fields that are not known analytically \cite{Sauvan2013, Bai2013, Vial2014}; 2) the completeness of QNM expansions inside and outside the resonators thanks to the incorporation of numerical modes in the expansion \cite{Vial2014, Yan2018}; and 3) the deployment of computational software \cite{Bai2013, Yan2018} that handle complicated 3D geometries. See \cite{Lalanne2018} for a recent review on QNMs, the definition of their mode volumes, quality factors, their various applications and the deeper physical insight that they convey into several important phenomena such as Purcell effect, strong coupling and cavity perturbation.
	
	In this work, we describe a novel approach based on QNM theory, which enables a deep analytical insight into the multi-parameter design required to optimize nonlinear nanophotonic structures. We firstly set the formalism framework, highlighting that once the nanoresonator eigenmodes are known, the linear and nonlinear responses are retrieved analytically. We then demonstrate the effectiveness of the QNM approach in terms of computational costs, design guidelines and simplicity of physical interpretation, by comparing the predictions of the formalism with exact data obtained with classical numerical solvers. Finally, we highlight the key outcome of the formalism: a closed-form expression, like for guided modes in integrated optics, of the complex overlap integral between the QNMs at the fundamental and harmonic frequencies. This leads us to propose a systematic design strategy to boost this overlap and enhance the efficiency of nonlinear processes in micro- and nano-scale resonators. Although in the following we will primarily focus on the concrete example of SHG, the proposed formalism can be extended to other second order, e.g. Sum/Difference Frequency Generation, and higher harmonics processes, as discussed in Section 3.
	\section{QNM theory of $\chi^{(2)}$ nanoresonators}
	\label{sec:theory}
	
	To set the formalism framework, we first consider an unsophisticated structure: a tiny resonator composed a material with a high nonlinear susceptibility tensor $\chi^{(2)}$, an AlGaAs nanocylinder, on a low index substrate, see Fig.~1a. This structure was considered in the first experimental demonstration of SHG with non-plasmonic nanostructures \cite{Gili2016}. Let us assume that SHG operates in the small-signal regime, where the lack of pump depletion leads to the well-known quadratic scaling of harmonic output with incident power \cite{Gili2016}. SHG can then be described via two coherent processes. An external driving field $[\textbf{E}_b (\textbf{r},\omega),\textbf{H}_b (\textbf{r},\omega)]$ first excites the resonator to generate a total field distribution $[\textbf{E}_t (\textbf{r},\omega),\textbf{H}_t (\textbf{r},\omega)]$ at the fundamental frequency (FF) $\omega$. We use a scattering-field formulation throughout the manuscript, see Annex 2 in \cite{Lalanne2018}, so that the driving field $\textbf{E}_b (\textbf{r},\omega)$  is composed of an incident plane wave with an electric field $\textbf{E}_0$ and a specularly reflected  plane wave with an amplitude fixed by the Fresnel reflection coefficient of the air-substrate interface. In a second step, the total FF field generates a local nonlinear current inside the resonator, $\textbf{J}^{(2)} (\textbf{r},2\omega)$, which acts as the source for the second harmonic (SH) radiation at $2\omega$. Hereafter, the incident plane wave is normalized such that its intensity is $S_0=1 GW/cm^2$.
	
	\begin{figure}[htbp]
		\centering
		\includegraphics[width=\linewidth]{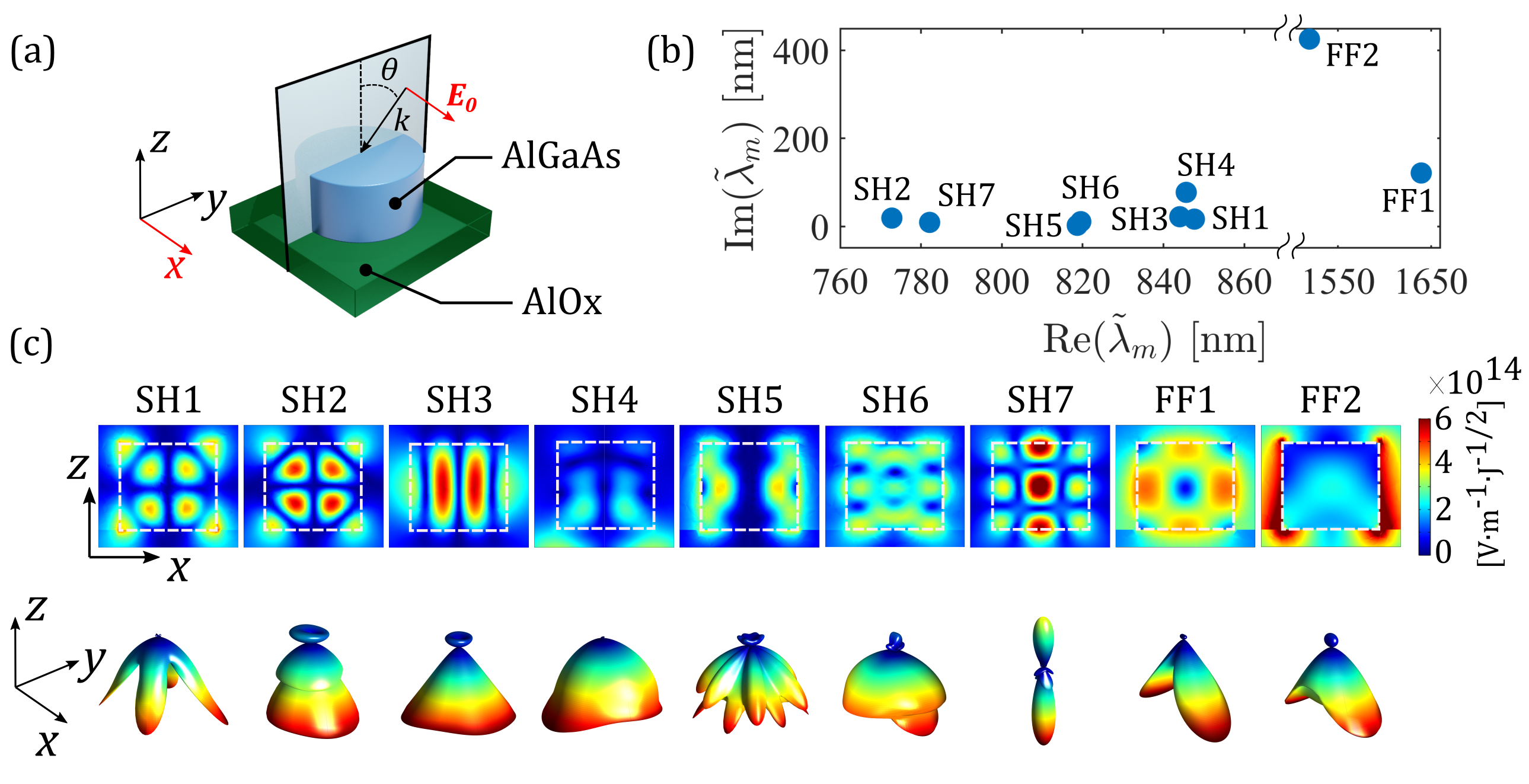}
		\caption{Set of QNMs excited at normal incidence for a simple geometry sketched in (a): an AlGaAs nanocylinder (radius $220$~nm, height $400$~nm) on an AlOx substrate. (b) Complex eigenfrequency of the main QNM near SH and FF. (c) Top: Corresponding electric field norm in the $xz$-plane at $y=0$. QNM near fields are normalized as in \cite{Lalanne2018}. Bottom: Radiation pattern for all the 9 QNM reported in (b). The refractive index of AlOx is 1.6. The AlGaAs Drude-Lorentz model parameters are given by $\varepsilon_\infty=1$, $\omega_p=1.69\cdot10^{16}rad/s$, $\omega_0=5.55\cdot10^{15}rad/s$, and $\gamma=0$ in the transparency window of AlGaAs ($\lambda>760$~nm). The COMSOL model used to obtain the figure can be downloaded with the QNMEig software\cite{Yan2018}.}
		\label{fig:fig1}
	\end{figure}
	
	QNM theory provides an ideal platform to model these processes because they naturally rely on the natural resonances at the fundamental and second-harmonic frequencies. Let us label by $m$ the QNM set that covers the large spectral range from $\omega$  to $2\omega$, and let us denote by $[\tilde{\textbf{E}}_m (\textbf{r}),\tilde{\textbf{H}}_m (\textbf{r})]$ the normalized electric and magnetic field distributions of the $m^{th}$ QNM, with complex frequency $\tilde{\omega}_m$ and quality factor $Q_m=-Re(\tilde{\omega}_m)/2 Im(\tilde{\omega}_m)$ (we use the $exp(-i\omega t)$ convention). 
	To make it more concrete, for the considered structure, we show the frequency positions in the complex plane of the dominant modes in Fig.~1(b) and their field distributions in Fig.~1(c). Importantly, we normalize the QNM fields such that $ \int[\tilde{\textbf{E}}_m \cdot (\partial\tilde{\omega}\pmb{\varepsilon}/ \partial\tilde{\omega}) \tilde{\textbf{E}}_m - \tilde{\textbf{H}}_m\cdot (\partial \tilde{\omega}\pmb{\mu}/\partial\tilde{\omega})\tilde{\textbf{H}}_m]d^3\textbf{r} = 1 
	$ \cite{Lalanne2018}. Because the QNMs are leaky modes, their field exponentially diverges away from the resonator in space and the computation of the integral requires some care. In this work, we have indifferently used the QNM solvers QNMEig \cite{Yan2018} or QNMPole \cite{Bai2013} of the free software package MAN to normalize the QNMs and to reconstruct the scattered fields in the QNM basis.
	
	The following formulation relies on a recent QNM auxiliary-field formalism \cite{Yan2018} particularly effective for analyzing resonators with dispersive materials and incorporates our latest improvements \cite{Wu2019}, which enhances the accuracy and convergence rate of QNM expansions that are necessarily truncated for numerical purposes. In that respect, we assume that the nanocylinder relative permittivity can be modeled with a single-pole Lorentzian function $\varepsilon(\omega)=\varepsilon_\infty-\varepsilon_\infty\omega_p^2/(\omega^2-\omega_0^2+i\omega\gamma)$, where $\varepsilon_\infty$, $\omega_p$, $\omega_0$ and $\gamma$ are fitted to empirical models \cite{Gehrsitz2000}. The formalism can be indifferently applied to multipole expansions. Combining the results in \cite{Yan2018,Wu2019}, we reconstruct the total field inside the resonator at $\omega$ as
	\begin{equation}
	\textbf{E}_t(\textbf{r},\omega) = \sum_{m=1}^{M_1}\alpha_m^{(1)}(\omega)\frac{\varepsilon(\Tilde{\omega}_m)-\varepsilon_\infty}{\varepsilon(\omega)-\varepsilon_\infty}\Tilde{\textbf{E}}_m(\textbf{r})
	\label{eq:rec_totalfield}
	\end{equation}
	where $\alpha_m^{(1)}(\omega)=\int_V\{[\varepsilon(\Tilde{\omega}_m)-\varepsilon_b]\Tilde{\omega}_m/(\Tilde{\omega}_m-\omega)+(\varepsilon_b-\varepsilon_\infty) \}\tilde{\textbf{E}}_m(\textbf{r})\cdot\textbf{E}_b(\textbf{r},\omega)d^3\textbf{r}$ is the modal excitation coefficient of the $m^{th}$  QNM \cite{Yan2018} at FF. Note that the integral is performed of the volume $V$ that defines the resonator in the scattered field formulation.
	The total field of Eq.~(1) generates a nonlinear displacement current in the resonator,
	\begin{equation}
	\begin{split}
	\textbf{J}^{(2)}(\textbf{r},2\omega) = & -i2\omega\textbf{P}^{(2)}(\textbf{r},2\omega)=\\
	& =-i2\omega\varepsilon_0\pmb{\chi}^{(2)}(2\omega,\omega,\omega):[\textbf{E}_t(\textbf{r},\omega)\otimes\textbf{E}_t(\textbf{r},\omega)]\\
	\end{split}
	\label{eq:rec_current}
	\end{equation}
	which acts as a source at $2\omega$ for the nonlinear radiation. $\otimes$ and $:$ notations stand for tensorial and contructed product respectively. The total field $\textbf{E}_t(\textbf{r},2\omega)$  at $2\omega$ can also be expanded in the QNM basis
	\begin{equation}
	\textbf{E}_t(\textbf{r},2\omega)=\sum_{m=1}^{M_2}\alpha_m^{(2)}(2\omega)\tilde{\textbf{E}}_m(\textbf{r})
	\label{eq:rec_totalfieldSH}
	\end{equation}
	with modal excitation coefficients $\alpha_m^{(2)}(2\omega)=-2\omega/(\tilde{\omega}_m-2\omega)\int_V\textbf{P}^{(2)}(\textbf{r},2\omega)\cdot\tilde{\textbf{E}}_m(\textbf{r})d^3\textbf{r}$ \cite{Yan2018}. Injecting the first expansion at FF, Eq.~(1), into Eq.~(2) and then into Eq.~(3), it is straightforward to derive a closed-form expression for the modal excitation coefficient at SH
	\begin{equation}
	\alpha_l^{(2)}(2\omega)=\sum_{m,n}\frac{-2\omega^3\zeta_{lmn}\xi_{mn}(\omega)}{(\tilde{\omega}_l-2\omega)(\tilde{\omega}_m-\omega)(\tilde{\omega}_n-\omega)}
	\label{eq:coefficientsSH}
	\end{equation}
	
	with
	\begin{subequations}
		\begin{equation}
		\xi_{mn}(\omega)=\frac{[\varepsilon(\tilde{\omega}_m)-\varepsilon_\infty][\varepsilon(\tilde{\omega}_n)-\varepsilon_\infty]}{[\varepsilon(\omega)-\varepsilon_\infty]^2}\alpha_m^{(1)}(\omega)\alpha_n^{(1)}(\omega)
		\end{equation}
		\begin{equation}
		\zeta_{lmn}=\varepsilon_0\int_V\tilde{\textbf{E}}_l(\textbf{r})\cdot \{\pmb{\chi}^{(2)}:[\tilde{\textbf{E}}_m(\textbf{r})\otimes\tilde{\textbf{E}}_n(\textbf{r})]\}d^3\textbf{r}
		\end{equation}
		\label{eq:csizeta}
	\end{subequations}
	
	The possibility to reconstruct the SH field with a closed-form expression involving only a few resonances is the key outcome of the present work. Notably, the analyticity of Eq.~(4) suggests that the design of nanoresonators with targeted nonlinear response may be performed with a few simulations at complex frequencies without resorting to series of real-frequency simulations. This will be demonstrated below. From the knowledge of the field distribution at $2\omega$ in the nanoresonator, many important quantities can be straightforwardly computed. If the driving field at $\omega$ is a plane wave, we may also compute the nonlinear extinction cross section $\sigma_{ext}^{(2)}(2\omega)$, a classical figure of merit defined as the ratio between the generated power at $2\omega$ and the intensity $S_0$ of the incident field \cite{Bai2013},
	\begin{equation}
	\sigma_{ext}^{(2)}(2\omega)=-\frac{\omega}{S_0}\int_V Im[\sum_{l=1}^{M_2}\alpha_l^{(2)}(2\omega)\tilde{\textbf{E}}_l(\textbf{r})\cdot\textbf{P}^{(2)^*}(\textbf{r},2\omega)]d^3\textbf{r}    
	\end{equation}
	
	Since we are considering a second-order nonlinear process, it is important to recall that $\sigma_{ext}^{(2)}$ scales linearly with the incident power. Remarkably, Eq.~(6) allows to separately study the contribution from different modes to the extinction at $2\omega$. More generally, Eqs. (4) and (5) simply highlight the physics of SHG in this nanoantenna, and they deserve a few important comments:
	
	\begin{itemize}
		\item Equation (4) tells us that the excitation of the $l^{th}$ QNM at $2\omega$ is effective only if two QNMs labelled $m$ and $n$ are efficiently excited at $\omega$ by the driving field ($\xi_{mn}(\omega)$ term), and if a good spatial overlap between FF and SH modes ($\zeta_{lmn}$  term) is ensured. In this respect, it is interesting to consider what happens if the three interacting QNMs are exactly matched with the FF and SH frequencies. Setting $Re(\tilde{\omega}_l)=2\omega$ and $Re(\tilde{\omega}_m)=Re(\tilde{\omega}_n)=\omega$ , one obtains a simplified expression for the modal excitation coefficient at SH, $\alpha_l^{(2)}(2\omega)=8i\sum_{mn}Q_lQ_mQ_n\zeta_{lmn}\xi_{mn}(\omega)$, thus retrieving that nonlinear interactions are enhanced by resonators that confine light for long times (high Q factors).
		
		\item Equation (5a) highlights the excitation by the driving field at $\omega$. Once the QNMs are known by computation, the modal excitation coefficients $\alpha_m^{(1)}$  and $\alpha_n^{(1)}$ as well as the spectral response of the nanoresonator at $\omega$ are known analytically for any driving field. The analyticity has important consequences, as it not only clarifies the role of the selective excitation of some resonances at $\omega$, but may also help engineering the shape of the incident beam for optimizing the efficiency of nonlinear conversion or harnessing nonlinear optical effects, as was very recently reported with plasmonic oligomers and cylindrical vector beams to dynamically tune the SHG \cite{Bautista2018}.
		
		\item 	Equation (5b) provides an analytic expression for the complicated spatial overlap integral $\zeta_{lmn}$ between the nonlinearly interacting modes, thereby opening a new path towards a thorough engineering of nanoresonator structures with high conversion efficiencies, as we will illustrate in Section 4. The analytic expression is likely to be the most important outcome of the present formalism. With the exception of recent theoretical works \cite{Rodriguez2007,Lin2016}, exact expressions of the overlap integral have not been explicitly clarified in earlier works. However, the coupled-mode formalism used in \cite{Rodriguez2007,Lin2016} relies on Hermitian theory and is valid only for closed systems without dissipation; in sharp contrast, the present formalism is valid for non-Hermitian open systems. This difference clearly emerges when comparing our Eq.~(5b) with Eq.~(3) in \cite{Lin2016}. While the latter formula involves complex conjugate values of the electric fields, both in the overlap integral with a triple $\tilde{\emph{E}}_l^*\tilde{\emph{E}}_m\tilde{\emph{E}}_n$ product and in the mode normalization with integrals of $\tilde{\textbf{E}}\cdot\tilde{\textbf{E}}^*$ products, no complex conjugation occurs in the expression of $\zeta_{lmn}$ in Eq.~(5b). Indeed, for the nearly-Hermitian high-Q modes of photonic-crystal cavities, the QNM electric fields are almost real, $Im(\tilde{E})/Re(\tilde{E})\propto O(Q^{-1})$ \cite{Lalanne2018}, and both approaches become identical. However, in general, Eq.~(3) in \cite{Lin2016} and Eq.~(5b) herein provide significantly different predictions for nanoresonators that support strongly localized resonances. Actually, for resonances with a significant leakage, the phases of every QNM-field components vary spatially in a complicated manner, and the products $\tilde{\emph{E}}_l^*\tilde{\emph{E}}_m\tilde{\emph{E}}_n$ and $\tilde{\emph{E}}_l\tilde{\emph{E}}_m\tilde{\emph{E}}_n$ promoted by Eq.~(3) in \cite{Lin2016} and Eq.~(5b) significantly differ. In addition, let us recall that the normalization based on $\tilde{\textbf{E}}\cdot\tilde{\textbf{E}}^*$ products is just incorrect \cite{Lalanne2018}. An in-depth analysis of the problems encountered when using Hermitian theory for open nanoresonators has been recently presented in the context of cavity perturbation theory \cite{Yang2015}.
		
		\item 	We emphasize that the spatial overlap-integral $\zeta_{lmn}$ quantitatively estimates the conversion efficiency between $\omega$ and $2\omega$. Since all the QNM fields are normalized in a unique manner, $\zeta_{lmn}$ is an intrinsic quantity that solely depends on them. There is no undetermined proportionality factor as in earlier works, and $|\zeta_{lmn} |^2$ strictly represents the conversion efficiency. Some illustrative values of $\zeta_{lmn}$ will be provided in section 4 for different resonant contributions. 
	\end{itemize}
	
	\section{Computational force of QNMs for nonlinear nano-optics modeling}
	
	In this section, we methodically present the different computational steps for implementing the QNM theory, considering the simple example of the AlGaAs-on-AlOx nanocylinder. We also restrict ourselves to pump wavelengths varying from $1600$~nm to $1750$~nm and pump incidence angles from $0^\circ$ to $50^\circ$. We expect to draw the reader attention on the simplicity of the implementation, highlighting the potential of the approach. The QNM-formulation predictions are systematically compared with exact numerical results obtained with COMSOL Multiphysics. We will refer to these reference data as “exact” data. Since we use the same finite-element fine mesh and the same workstation to compute the QNMs and the exact data, the computational accuracy and CPU times can be fairly compared.
	
	We start by computing the relevant QNMs with QNMEig. In order to optimize computation times, we restrict the pole search to our two spectral regions of interest, finding 10 modes around the FF wavelength $\lambda =1650$~nm and 30 modes around the SH wavelength $\lambda = 825$~nm. This computation requires around 5 minutes on a workstation; it also represents the only numerical computation since the formalism provides analytical expressions for the field reconstruction at the FF and SH frequencies.
	\begin{figure*}[htbp]
		\centering
		\includegraphics[width=\textwidth]{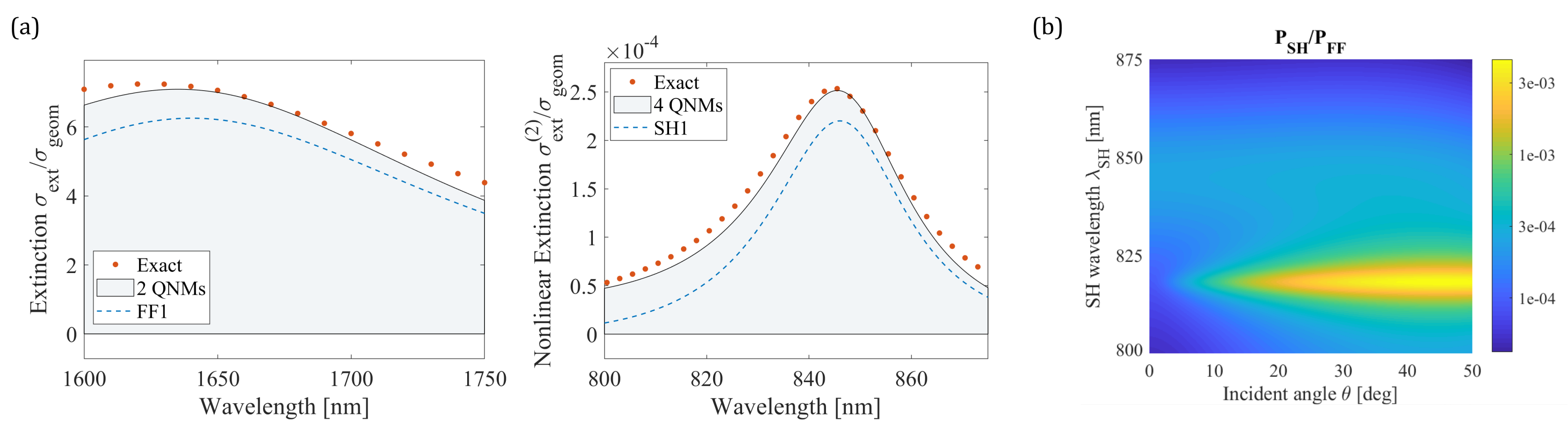}
		\caption{SHG by an AlGaAs-on-AlOx nanocylinder with $r =~220$~nm and $h = 400$~nm. (a) Linear (left) and SHG (right) extinction efficiencies. Approximate cross sections (black lines) reconstructed with only a few most relevant QNMs (FF1-FF2 and SH1-SH4), shown in Fig.~1, compared with fully vectorial numerical results (red bullets). Blue dashed lines show the contribution just a single QNM. For the computations, we assume that the nanocylinder is illuminated by a plane wave with linear polarization along $x$ and $k$ vector along $z$. (b) Ratio between SHG and FF power versus incidence angle $\theta$ and SH wavelength (the latter corresponding to $\lambda_{FF}$  spanning from $1600$ to $1750$~nm).}
		\label{fig:fig2}
	\end{figure*}
	The present work being solely intended to evidence the potential of the QNM formalism for nonlinear studies and designs in nanophotonics, we considerably reduce this initial set, considering only $M_1=2$ QNMs at FF and $M_2=7$ QNMs at SH. The interested reader may refer to the Supplementary Information in \cite{Yan2018} for a careful analysis of the convergence performance of QNM expansions.
	
	The electric-field norm of these dominant QNMs are shown in Fig.~1(c), along with their radiation diagrams. The later provide valuable information on the QNM excitation probability at FF, or on the QNM contribution to the far-field pattern at $2\omega$. Since the near-to-far field transform of COMSOL-Multiphysics is only valid for scatterers in a uniform background, we have used the freeware RETOP \cite{Yang2016} to calculate the radiation diagrams in the air and AlOx clads. Additionnally note that RETOP does not handle complex frequencies and the transformations are approximately performed at the real frequencies of every QNM.
	
	Once the modes $[\tilde{\textbf{E}}_m,\tilde{\textbf{H}}_m]$ are known, we analytically compute the excitation coefficients $\alpha_m^{(1)} (\omega)$ at FF using the toolboxes provided in QNMEig and then reconstruct the total field $\textbf{E}_t (\textbf{r},\omega)$ inside the resonator at FF. From the knowledge of the total field, many important physical quantities are derived. The linear extinction spectrum, computed as in the Supplementary information in \cite{Yan2018}, is reported in the left panel of Fig.~2(a) for normal incidence, and compared with exact data directly obtained with COMSOL for every frequency. A quantitative agreement, even when a very small number ($M_1=2$) of QNMs is retained in the expansion of Eq.~(1), is achieved.
	Then, the nonlinear displacement currents are straightforwardly obtained with Eq.~(2). We further compute the modal excitation coefficients $\alpha_m^{(2)} (2\omega)$ and reconstruct the total field $\textbf{E}_t (\textbf{r},2\omega)$  with Eq.~(3). These computations are similar to those performed at FF. The predicted nonlinear extinction cross section $\sigma_{ext}^{(2)} (2\omega)$, obtained with Eq.~(6), is shown in the right panel of Fig.~2(a). Again, a quantitative agreement with exact data is achieved over the entire spectrum, even when a very small number ($M_2=4$) of QNMs is retained in the expansion of Eq.~(3). Advanced details concerning the numerical implementation of the method can be found in the QNMEig workpackage freeware. These include a released COMSOL model sheet of the AlGaAs-on-AlOx nanocylinder and the companion Matlab script used to compute the $\alpha_m^{(2)} (2\omega)$ and $\alpha_m^{(1)} (\omega)$ and reconstruct the scattered fields.
	
	In Fig.~2(b), we plot the SHG extinction efficiency $P_{SH}/P_{FF}$ as a function of the SH wavelength and the incidence angle $\theta$ of the FF plane wave. $P_{SH}/P_{FF}$ is defined by the ratio of the SHG power to the FF power $S_0 \pi r^2$ impinging on the nanocylinder. The results are all obtained for a linearly transverse-polarized plane wave parallel to the $x$-axis. For the chosen spectral and angular resolutions, 7500 different instances for the background field have been explored. Based on Eqs. (1)-(3), the whole SHG efficiency map is computed in only 2 minutes (this CPU time does not include the initial computation of the QNMs made once for all). By contrast, since the estimated time for a single fully numerical simulation in COMSOL-Multiphysics on the same machine is $\approx$ 2 minutes, the calculation of the same map would require $\approx$ 10 days.
	
	The CPU time reduction is by no means the only positive aspect of the method. The possibility to directly assess the contribution of every individual QNM to the SH extinction, thereby allowing to identify the dominant modes. For instance, the knowledge of the normalized near-field distributions of the four dominant SH modes gives a deep insight into the nonlinear conversion occurring inside the nanocylinder (this will be analyzed in the next Section). The radiation pattern of every individual QNM excited at SH additionally gives valuable information on the spatial directions for which the SH signal can be effectively observed in the far field. From the symmetries of the dominant modes at FF, the best ways to tailor the polarization of the pump beam and to selectively favor the excitation of a resonant mode \cite{Carletti2016, Carletti2018} and to control the nonlinear process can be quantitatively analyzed.
	Finally, note that rewriting $\textbf{J}^{(2)}$ in Eq.~(2) for a non-degenerate $\chi^{(2)}$  process, we can straightforwardly retrieve the versions of Eqs. (4-6) for sum/difference frequency generation and parametric down-conversion. Similarly, substituting $\textbf{J}^{(2)}$ in Eq.~(2) with third-order nonlinear current $\textbf{J}^{(3)}$, the entire model can be generalized to $\chi^{(3)}$ processes, thereby describing other important effects like third-harmonic generation, four-wave mixing, self-phase modulation, and cross-phase modulation. While formal changes to Eqs. (4-6) are quite trivial for generalization to $\chi^{(3)}$ and higher order $n$ of the nonlinearity $\chi^{(n)}$, the number of QMNs to be considered significantly grows with $n$, implying that the simplicity and transparency of the nonlinear QNM formalism may become questionable for high-order nonlinearities.
	
	\section{Importance of mode matching for boosting nonlinear conversion}
	
	The quasinormal-mode formalism developed in Section 2 enables a deep analytical insight into the complicated multi-spectral harmonic conversion occurring inside $\chi^{(2)}$ nanoresonators. In this Section, we explore new paths that use this insight to optimize the nonlinear generation at subwavelength scale. The approach relies on the knowledge of the normalized near-field distributions of a few dominant modes and on the closed-form expressions governing the nanoresonator response at $\omega$ on the one hand, and the complicated spatial overlap integral $\zeta_{lmn}$ between the nonlinearly interacting modes at FF and SH on the other.
	
	For the sake of illustration, we again consider an AlGaAs-on-AlOx nanocylinder drilled by an axial hole with an rounded-rectangle cross-section, see the inset in Fig.~3(b). Thanks to the hole, the degeneracy of the axisymmetric modes is lifted and additional resonances are revealed. A similar effect may be obtained with elliptical nanocylinders, our choice being motivated by fabrication issues in relation with the $\chi^{(2)}$ engineering approach presented below. The cylinder is assumed to be driven by a Gaussian pump beam (beam waist $w_0=2 \mu m$) that is normally incident. The initial step of the design, not reported for the sake of compactness, has consisted in optimizing the nanocylinder dimensions to guaranty that the nanocylinder supports at least one high-Q resonance at the SH (we target a SH wavelength of $800$~nm). After a few QNM computations, we have selected a nanocylinder with slightly larger dimensions than in Fig.~1 ($440$~nm radius and $400$~nm height), and hole sizes $l=480$~nm and $w=150$~nm, offering a resonance (labelled FF1) at FF with a central wavelength of $1674$~nm and a quality factor of 8.4. The nanocylinder response at FF is largely dominated by the excitation of this mode. Its electric field distribution is reported in Fig.~3(a), along with those of seven other relevant QNMs.
	
	\begin{figure}[h]
		\centering
		\includegraphics[width=\linewidth]{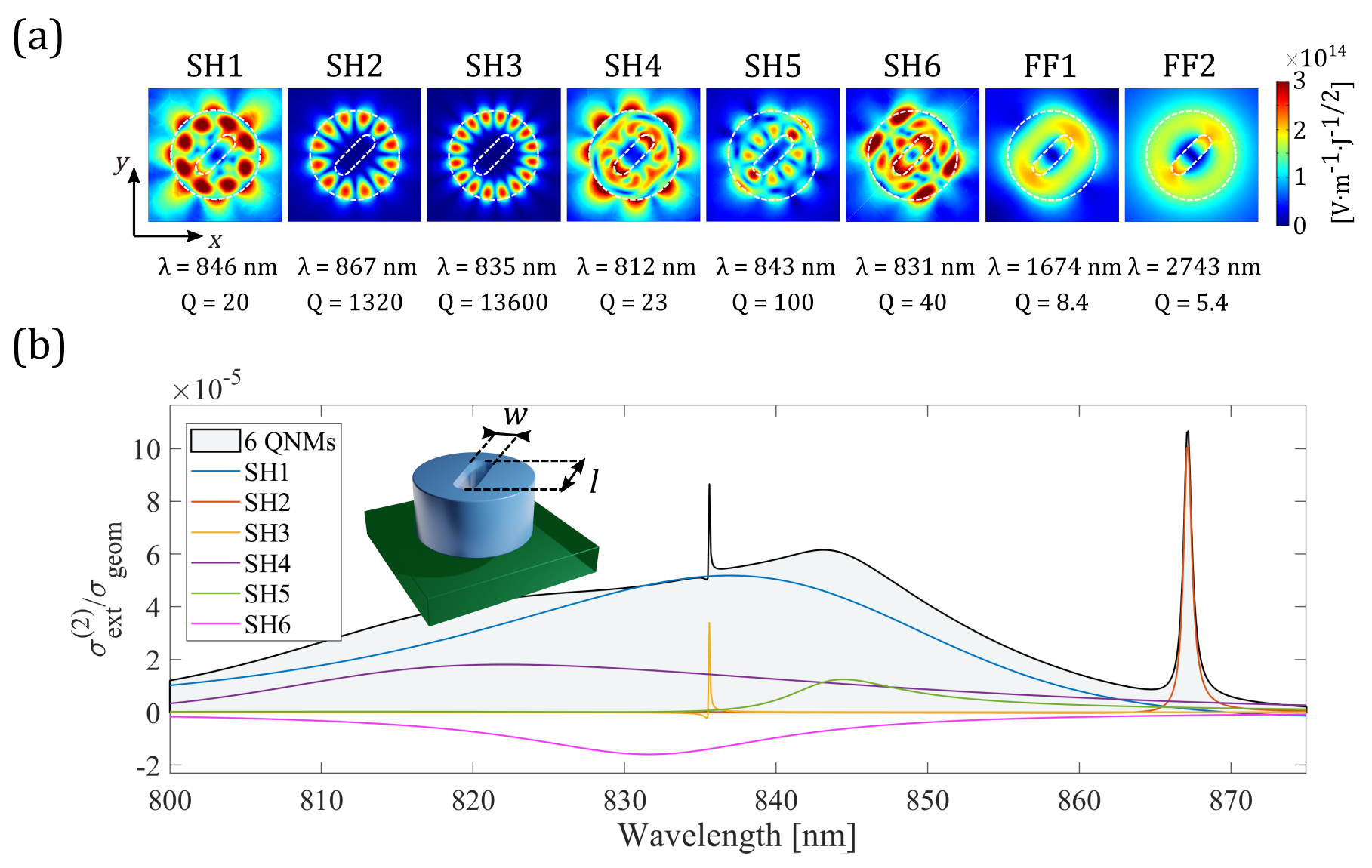}
		\caption{SH scattering by an AlGaAs-on-AlOx cylindrical nanostructure with radius $r = 440$~nm, height $h = 400$~nm and a central asymmetric hole with length $l = 480$~nm, width $w = 150$~nm and deep all the resonator height. (a) Electric-field norm in $xy$-plane of two QNMs around FF and six around SH. (b) Contribution to the SH extinction cross sections from the six most excited QNMs around SH frequency reported in (a), normalized by the geometric cross section.}
		\label{fig:fig3}
	\end{figure}
	
	\subsection{Mode matching at FF}
	
	The closed-form expression of the modal excitation coefficients, $\alpha_m^{(1)} (\omega)$, suggests that the integral of the scalar product $\tilde{\textbf{E}}_m(\textbf{r})\cdot\textbf{E}_b(\textbf{r},\omega)$ in the resonator volume has to be maximized to efficiently couple the FF beam with a specific mode, in addition to matching the pump frequency and the resonance frequency. Since the electric-field distribution of the FF1 QNM has a prevailing azimuthal polarization in the $xy$-plane, a gaussian azimuthally polarized beam impinging at normal incidence from air appears to be the most natural choice to pump the resonator. Thus, the incident field $\textbf{E}_{inc}$ close to the interface can be approximated in cylindrical coordinates as $\textbf{E}_{inc}(\rho,\phi,z) = \frac{A}{\sqrt{\pi}}\frac{2\rho}{w(z)^2} e^{-(\frac{\rho}{w(z)})^2(1-i\frac{z}{z_R})}e^{-2itan^{-1}(z/z_R)}e^{ikz}\hat{\pmb{\phi}}$ \cite{Veysi2016}, with $w(z)=w_0\sqrt{1+(z/z_R)^2}$ the beam radius, $z_R=\pi w_0^2/\lambda$  the Rayleigh range, $k=2\pi/\lambda$ the wavenumber, $A$ an amplitude coefficient (in Volts) and $\hat{\pmb{\phi}}$ the azimuthal unitary vector. Hereafter we choose a $50 W$ power for the driving field at $\omega$, a reasonable value for typical laser pulses in nonlinear nanophotonics \cite{Camacho-Morales2016}. Since the FF beam is not a plane wave, we normalize the nonlinear extinction cross section in Eq.~(6) by the spatially averaged power incident on the nanocylinder.
	
	\subsection{Mode matching of nonlinearly interacting modes}
	
	Six QNMs, labelled with the subscripts $l (l=1,…6)$, with resonance wavelengths $\tilde{\lambda}_l$ around $800$~nm are dominantly excited during the nonlinear conversion. Table 1 reports the values of their quality factor $Q_l$. The individual contributions of the six QNMs to the generated SH signal are calculated with Eq.~(6) and are shown in Fig.~3(b). One of them is dominantly negative, implying that the corresponding QNM (SH6) detrimentally contributes to the SH generation. This effect is similar to the “negative Purcell effect” reported in \cite{Sauvan2013} and occurs whenever QNMs spectrally overlap and interact. We additionally note that this has already been shown in literature \cite{Powell2017} for the linear extinction cross section of an air-suspended silicon nanodisk. Two QNMs, $l=2,3$ are whispering-gallery modes with large Q’s values and contribute to the SH generation in tiny spectral ranges.
		\begin{table}[h]
		\centering
		\caption{\bf Quality factors and spatial overlap integrals for the dominant SH modes}
		\begin{tabular}{ccccccc}
			\hline
			Mode & SH1 & SH2 & SH3 & SH4 & SH5 & SH6 \\
			\hline
			$Q_l$ & $20$ & $1320$ & $13600$ & $23$ & $100$ & $40$ \\
			$|\zeta_{l11}|$ & $151$ & $12$ & $11$ & $42$ & $29$ & $106$ \\
			\hline
		\end{tabular}
		\label{tab:overlap}
	\end{table}
	
	The spatial overlap-integral $\zeta_{lmn}$ is an important figure of merit of the present nonlinear QNM formalism. It is an intrinsic quantity, which is completely independent of the pump beam and may be skillfully used to quantify the nonlinear conversion efficiency. In Table 1, we provide the six values of $|\zeta_{l11}|$ associated to the FF1 mode. As expected from the absence of symmetry, none of the overlaps is null; however it is noticeable that their values significantly differ (remember that the QNMs are normalized and that the $|\zeta_{l11}|$ values can be compared with each other), implying that some QNMs naturally offer a good phase-matching. Unfortunately, these QNMs have low Qs. 
	Let us now illustrate how we may use the information brought by the spatial overlap integral to optimize the nonlinear conversion. We first select one of the six QNMs. Since the modal excitation coefficient at SH, $\alpha_l^{(2)} (2\omega)$, linearly scales with $Q_l$, we conveniently consider the SH mode with the highest quality factor, in this specific case SH3.

	\begin{figure}[h]
		\centering
		\includegraphics[width=\linewidth]{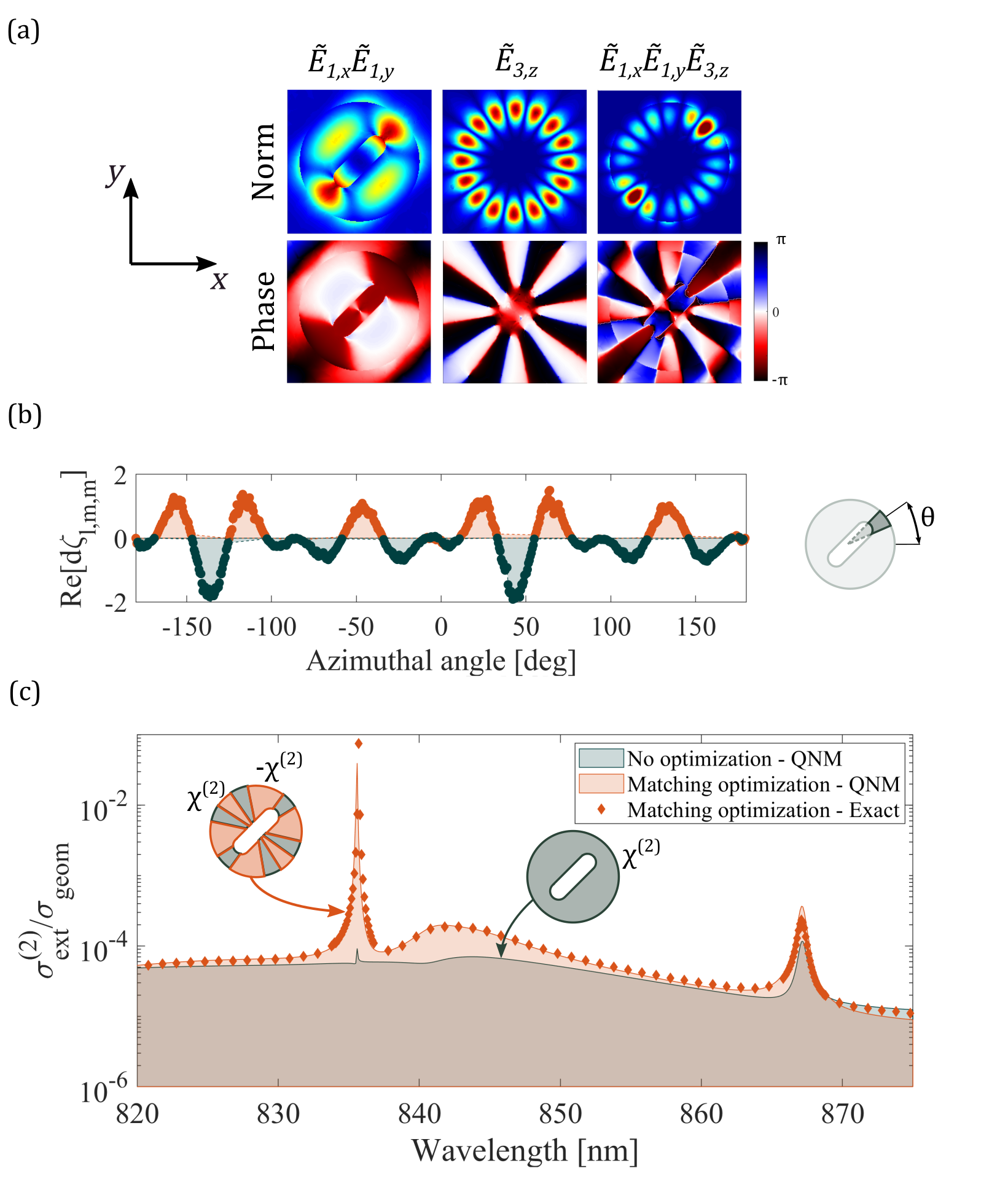}
		\caption{QNM-assisted SHG optimization for the nanoresonator of Fig.~3. (a)  Norm (top) and phase (bottom) of the electric near-field distribution in the $xy$ median plane for the main QNM excited at $\lambda_{FF}=1670$~nm ($\tilde{\textbf{E}}_m$) and the main QNM excited at SH ($\tilde{\textbf{E}}_l$). Left: product $\tilde{\emph{E}}_{m,x}\tilde{\emph{E}}_{m,y}$ of the two main components of the mode at FF. Center: distribution $\tilde{\emph{E}}_{l,z}$ of the most excited mode component at SH. Right: product $\tilde{\emph{E}}_{m,x}\tilde{\emph{E}}_{m,y}\tilde{\emph{E}}_{l,z}$ which is the main contribution to $\zeta_{lmn}$. (b) Discretization of $\zeta_{lmn}$, showing the contribution of $N$ azimuthal slices (see inset).  (c) SH to FF power ratio in the same conditions as in Fig.~3, for a homogeneous $(001)$ AlGaAs nanocylinder (grey line) and an optimized one where half of the slices have reversed $\chi^{(2)}$ orientation, i.e. $(00\bar{1})$  AlGaAs (orange line). Red dots show the full vectorial result computed in COMSOL for the case of optimized structure.}
		\label{fig:fig4}
	\end{figure} 
	
	FF1 being mainly polarized along the $x$- and $y$-directions, the integral in Eq.~(5b) is dominated by the term $\tilde{E}_{3,z}(\textbf{r})\tilde{E}_{1,x}(\textbf{r})\tilde{E}_{1,y}(\textbf{r}) $. In Fig.~4(a), we plot two-dimensional median cross-section maps of the module and phase of $\tilde{E}_{1,x}(\textbf{r})\tilde{E}_{1,y}(\textbf{r}) $ and $\tilde{E}_{3,z}(\textbf{r})$, therein providing a direct visual representation of the nanocylinder regions that have positive or negative contributions to the overlap integral, see the white and dark angular sectors in the bottom-right panel in Fig.~4(a). To be more quantitative, in Fig.~4(b), we show the overlap-integral averaged over radial planes of the nanocylinders as a function of the azimuthal angle. 
	
	In order to enhance the excitation coefficient and consequently boost SH generation, a possible solution is to locally reverse the sign of the $\chi^{(2)}$ tensor, while keeping the cylinder permittivity (and thus the QNMs) unchanged. Inspired by Fig.~4(b), we divide the nanocylinder in 12 angular sectors with opposite $\chi^{(2)}$, leading to the azimuthally-poled device shown in the inset of Fig.~4(c). The fabrication of such a poled cylinder, with different GaAs crystalline orientations in a subwavelength structure, represents a technological challenge. However, we note that similar devices have been recently fabricated by combining a single lithographic process with an epitaxial regrowth on a thin Ge adlayer and have successfully implemented quasi-phased matching in linearly-poled GaAs waveguides~\cite{Vodopyanov2004,Eyres2001}. The various contributions to $\zeta_{lmn}$ are then re-phased and optimal SHG by mode-matching is expected for the nanocylinder.
	
	In Fig.~4(c), we compare the nonlinear extinction spectra of the initial and optimized nanocylinders. The spectra are both reconstructed with the 8 QNMs shown in Fig.~3(a). Additionally, as a final evidence, we also provide the nonlinear extinction spectrum directly computed in the frequency domain with COMSOL-multiphysics. The numerical data shown with the red dots are in excellent agreement. Remarkably,  the SHG power is enhanced by more than two orders of magnitude for a pump at $\lambda_{FF} = 1670$~nm, highlighting the relevance of the spatial overlap integral $\zeta_{lmn}$ for design.
	
	\section{Conclusion}
	Nonlinear nanophotonics testified in recent years the emergence of a plethora of solutions to create novel sub-wavelength resonators with tailorable radiation properties. In most of the cases, all-dielectric nanoantenna design has been based on Mie-theory. Here we demonstrate that quasinormal mode expansion provides a precious theoretical formalism and numerical tool to model the nonlinear behavior of such open resonators. By combining a drastic reduction of computational costs with a deeper physical insight into the resonant behavior of dielectric nanoparticles, this method paves the way to a systematic and effective approach for the design of nonlinear subwavelength devices and the comprehension of their limits.
	
	\section*{Acknowledgments}
	
	The authors thank A. Gras and M. Ravaro for fruitful discussions. GL and PL acknowledge NOMOS project (ANR-18CE24-0026) for financial support.


\bibliography{biblio}
\end{document}